\documentclass[%
 reprint,
 amsmath,amssymb,
 aps,
refs,
]{revtex4-2}

\usepackage{graphicx}
\usepackage{dcolumn}
\usepackage{bm}
\usepackage{textcomp}
\usepackage{natbib}
\begin{document}

\preprint{APS/123-QED}

\title{Optimizing Beam-Plasma Interactions Through Jitter Analysis Using Start-to-End Simulations}

\author{R. Hwang}
\email{rlhwang@umich.edu}
\affiliation{SLAC National Accelerator Laboratory, Menlo Park, CA 94025, USA}

\date{August 20, 2024}

\begin{abstract}
Traditional accelerators, while effective, suffer from extensive spatial and financial demands, necessitating the exploration of compact alternatives like PWFA, which significantly reduces the necessary accelerator length by utilizing the wake generated by a high-speed pulse traveling through plasma. Our research focuses on mitigating instabilities, particularly timing jitter, which critically impacts the quality of accelerated beams. Through the deployment of Impact-T, Bmad, and Tao simulation tools at the FACET-II facility, we examined how timing jitter influences key beam parameters, including peak currents and emittance, over various simulation scenarios. The findings reveal that even minute variations in accelerator settings can significantly influence beam characteristics, underscoring the importance of precise control in beam dynamics. The outcomes contribute to enhancing the reliability and precision of PWFA systems, promising improved applications in both scientific research and medical therapies. Future research directions include integrating machine learning techniques to refine control strategies further and reduce experimental redundancies, highlighting the evolving synergy between accelerator physics and computational data science.
\end{abstract}

\maketitle


\section{Introduction}
Particle accelerators have long been central to advancements in both fundamental physics and applied sciences. Traditional particle accelerator systems, as illustrated in Figure 1, utilize radiofrequency (RF) waves to accelerate particles along extensive beamlines. For instance, FACET-II, a state-of-the-art accelerator, achieves an energy gain of 0 to 10 GeV over a length of 1 kilometer. However, such conventional accelerators face significant challenges, notably their large spatial footprint and associated costs. Even a relatively short 0.1-kilometer accelerator can require substantial financial investment, ranging from 10 million to 100 million (USD).

\begin{figure}[ht]
    \centering
    \includegraphics[width=\linewidth]{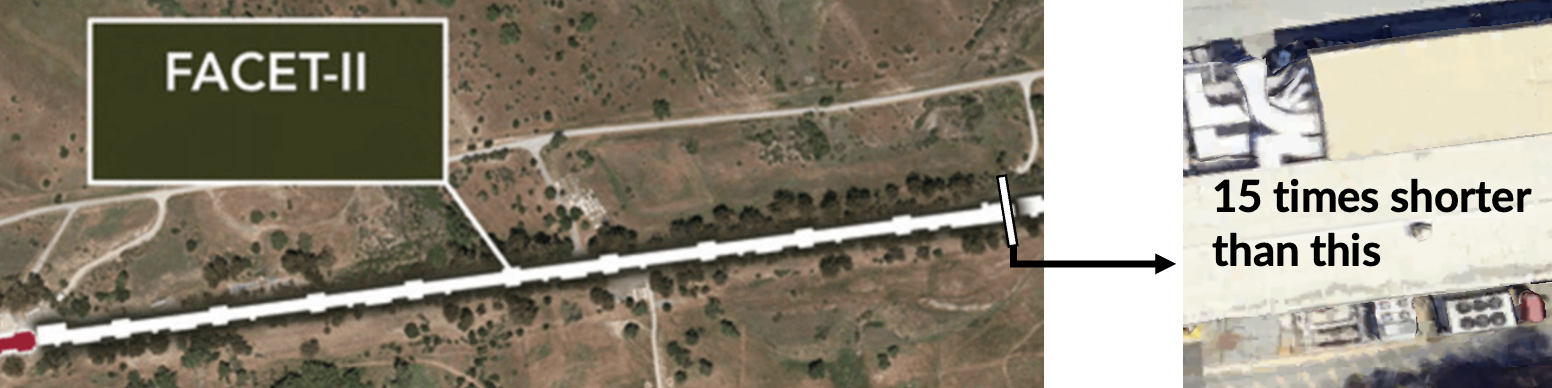}
    \caption{FACET-II Beamline vs. Experimental Area}
   \label{fig:synthetic_reconstruction}
\end{figure}

These challenges highlight the need for more compact and cost-effective accelerator technologies, which has led to increased interest in smaller particle accelerators. One prominent application of these smaller systems is in Proton-Ion Therapy, where synchrotrons are currently used. There is significant potential for developing smaller, more accessible medical accelerators, which could revolutionize cancer treatment by making advanced therapies more widely available.

The discovery of new particles, such as the Higgs Boson at the Large Hadron Collider, also underscores the critical role of accelerators in exploring fundamental physics. However, the pursuit of higher energies in smaller spaces brings about new challenges, particularly regarding the stability and quality of the particle beams. The intuitive approach of combining multiple plasma accelerators into a single large system encounters physical limitations, particularly the risk of electrical discharge damage at higher energies.

Plasma Wakefield Acceleration (PWFA) represents a promising alternative. PWFA has demonstrated the capability to achieve a 10 GeV energy gain within just 1 meter, a significant reduction in the required accelerator length compared to conventional methods. This technique utilizes the wake generated by a high-speed pulse traveling through plasma, akin to the wake produced by a boat moving through water. The energy within this wake can be harnessed to accelerate particles to high energies over much shorter distances.


Despite the advantages of PWFA, the quality of the accelerated beams remains a critical issue, particularly for high-precision experiments. Variations within the accelerator, such as timing jitter, can significantly impact beam quality. Understanding and mitigating these instabilities is essential for optimizing beam-plasma interactions, which is the primary focus of this research.

The proposed study aims to optimize beam-plasma interactions through detailed jitter analysis using start-to-end simulations. By examining how small instabilities, such as timing jitter, propagate and affect the final beam quality, this research seeks to develop methods to improve the performance of PWFA systems. The goal is to enhance the reliability and precision of these accelerators, making them more viable for both scientific research and practical applications like medical therapies.


\section{Simulation Setup}

The research conducted at FACET-II utilized advanced simulation tools to study the impact of jitter on beam quality. The workflow involved several steps, beginning with the selection of an optimal beam using the Impact-T simulation software, which simulates the injector portion of the accelerator. The selected beam was then used as input for further simulations using Bmad and Tao software on the SLAC Supercomputer Facility's (S3DF) Roma cluster. The Rome cluster is equiped with AMD EPYC Rome 120-core central processing units (CPUs).

The simulations were run for 200 iterations, with the parameters varied within one standard deviation of a Gaussian distribution. This approach allowed the researchers to study a broad range of scenarios and understand how small variations in the input parameters could affect the final beam quality. The parameters studied included the phase and amplitude of the L1 and L2 cavities and the source charge, with sigma values ranging from 0.3 to 2.3. Sigma values for each of the five parameters are outlined in figure 3 below.

    \begin{figure}[ht]
    \centering
    \includegraphics[width=0.6\linewidth]{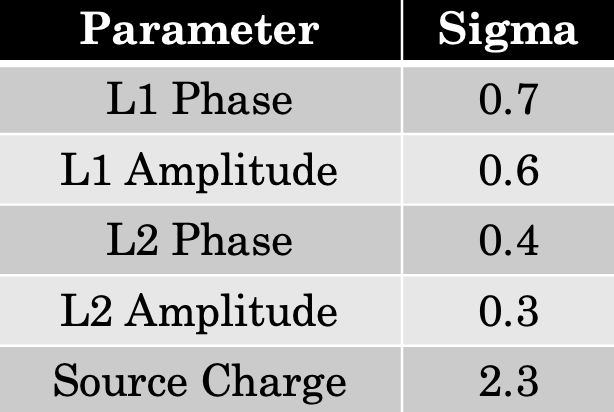}
    \caption{Parameter Sigma Values}
    \label{fig:synthetic_reconstruction}
\end{figure}

Data from the simulations were collected at four key locations along the beamline: the end of Bunch Compressors 11 and 14 (BC11 and BC14), the beginning of BC20, and the plasma entrance. The data were then compressed using a lossless algorithm and stored in a structured format for further analysis.

The Deflate algorithm, which was used in this research, is a widely used lossless data compression algorithm that combines the LZ77 algorithm and Huffman coding. LZ77 identifies and eliminates redundancy in the data by finding repeated sequences of bytes and replacing them with shorter references. For example, if a certain sequence of bytes occurs multiple times in the data, LZ77 will replace each occurrence with a pointer to the first occurrence, thereby reducing the size of the data. After LZ77 has compressed the data, Huffman coding is applied to further reduce the size. Huffman coding is a method of entropy encoding that assigns shorter codes to more frequent symbols and longer codes to less frequent symbols. This step is particularly effective at reducing the size of the already compressed data from the LZ77 stage.

By combining these two methods, the Deflate algorithm can achieve significant reductions in data size while ensuring that the original data can be perfectly reconstructed when decompressed. The ouputs were stored as specialized binary data in a custom file type called the .ROBIN file, which saw a three-fold decrease in file size compared to using UTF-8 encoded text.

\section{Results}

The jitter analysis focused on six key outputs: peak current, spot size in the x and y directions, emittance in the x and y directions, and the current profile. By analyzing these outputs, the research sought to identify the most significant sources of jitter and understand how they impact beam quality.

The peak current is a critical parameter for many applications, as it directly affects the beam's intensity and, consequently, its effectiveness in experiments or treatments. The simulations revealed that variations in the L1 and L2 phase and amplitude could significantly impact the peak current, with even small changes in phase (as little as one degree) leading to noticeable differences. However, the beam proved to be generally good at maintaining a consistent peak current from shot to shot, as shown in figure 4.

\begin{figure}[ht]
    \centering
    \includegraphics[width=\linewidth]{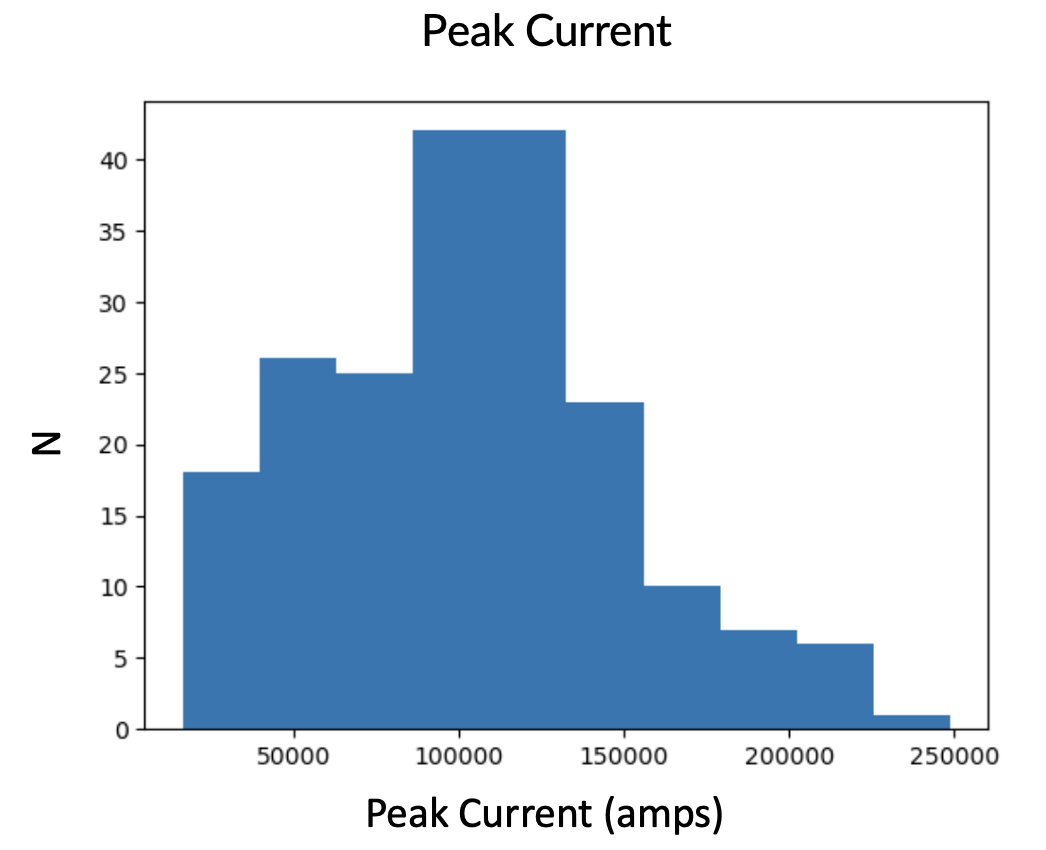}
    \caption{Distribution of Peak Currents Across 200 Simulations}
    \label{fig:synthetic_reconstruction}
\end{figure}

Similarly, the spot size, which refers to the beam's cross-sectional area, was found to be highly sensitive to jitter in the accelerating cavities. A smaller spot size is generally desirable as it indicates a more focused beam, which is essential for precision applications. The research showed that jitter in the L1 and L2 cavities could lead to fluctuations in the spot size, potentially degrading the beam's quality.

Emittance is a measure of the beam's spread in both position and momentum space. Low emittance is crucial for maintaining a high-quality beam, as it indicates that the particles are well-aligned and moving in the same direction. The simulations revealed that jitter in the source charge and the accelerating cavities could lead to increased emittance, thereby reducing the beam's overall quality.

The L2 phase in a particle accelerator plays a critical role in shaping the longitudinal phase space, which is a representation of the particle beam's distribution in terms of its position and momentum along the beamline. The phase of the L2 cavity directly influences the energy modulation and timing of the particle bunches as they travel through the accelerator. Small variations in the L2 phase can lead to significant changes in the energy distribution of the particles, thereby affecting the overall compression and focus of the beam. 

\begin{figure}[ht]
    \centering
    \includegraphics[width=\linewidth]{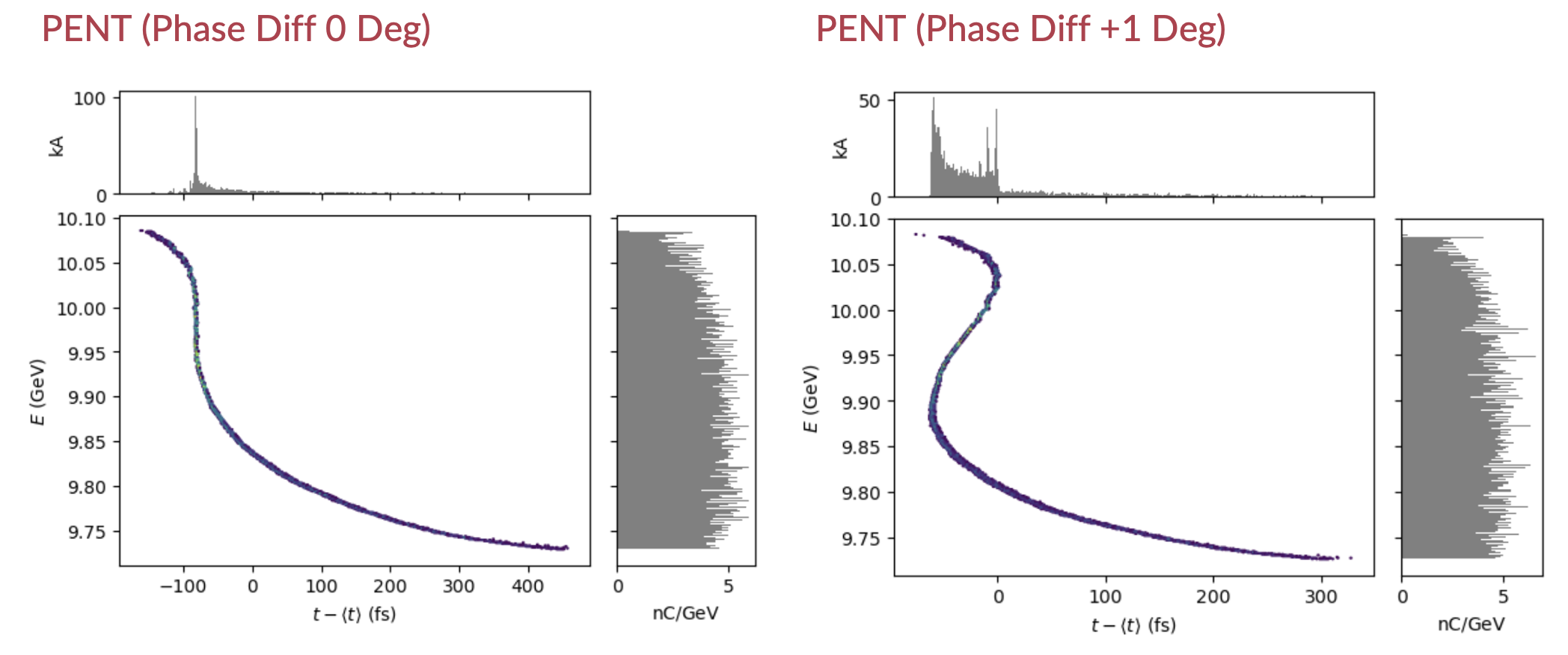}
    \label{fig:synthetic_reconstruction}
\end{figure}
\begin{figure}[ht]
    \centering
    \includegraphics[width=0.4\linewidth]{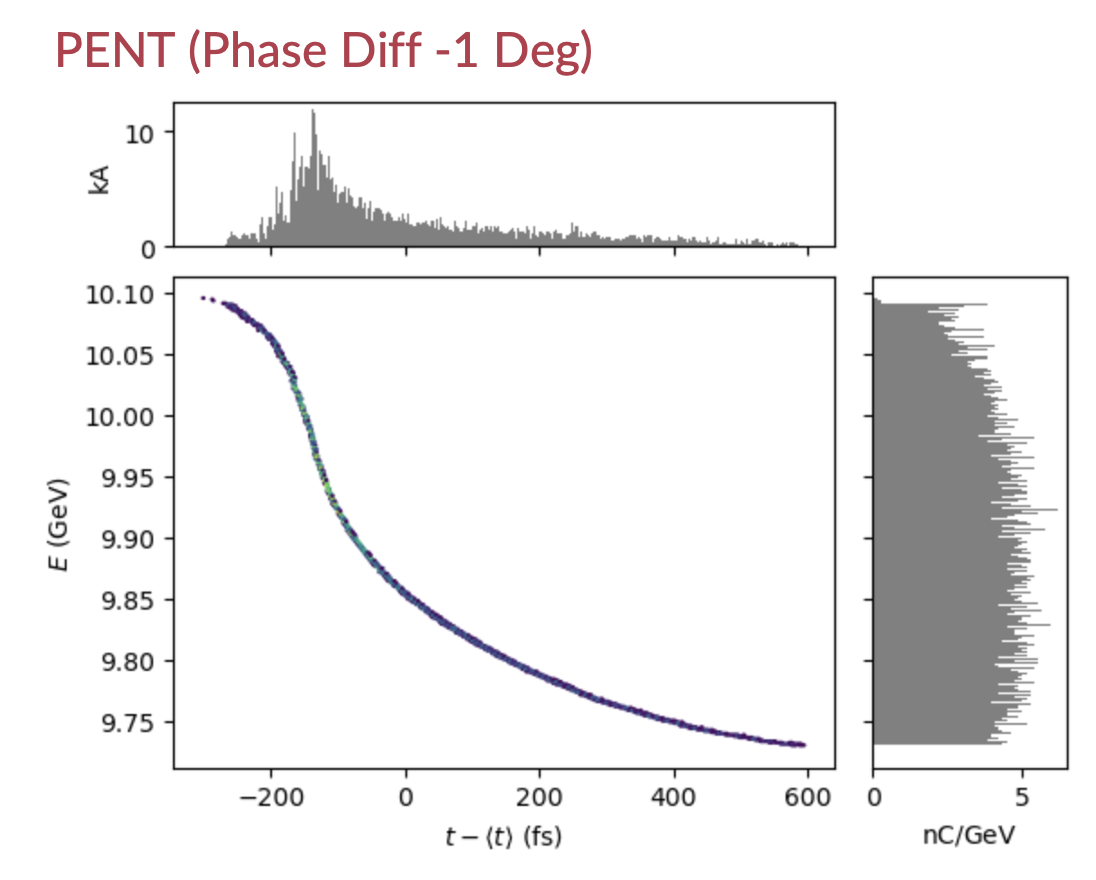}
    \caption{Longitudinal Phase Spaces with Optimal Beam vs. Beam with +/- 1 Degree of L2 Phase Jitter}
    \label{fig:synthetic_reconstruction}
\end{figure}

This, in turn, impacts the beam's longitudinal phase space, determining the quality and stability of the beam as it progresses through subsequent stages of acceleration. A well-optimized L2 phase is essential for maintaining a tight, synchronized bunch of particles, ensuring that the beam retains high quality with minimal energy spread, which is crucial for precision experiments and applications that rely on high beam fidelity. Figure 5 demonstrates the visual and quantitative differences in the longitudinal phase space between two phase values, thus proving how even 1 degree of jitter applied to phase can significantly impact the beam.

\begin{figure}[ht]
    \centering
    \includegraphics[width=\linewidth]{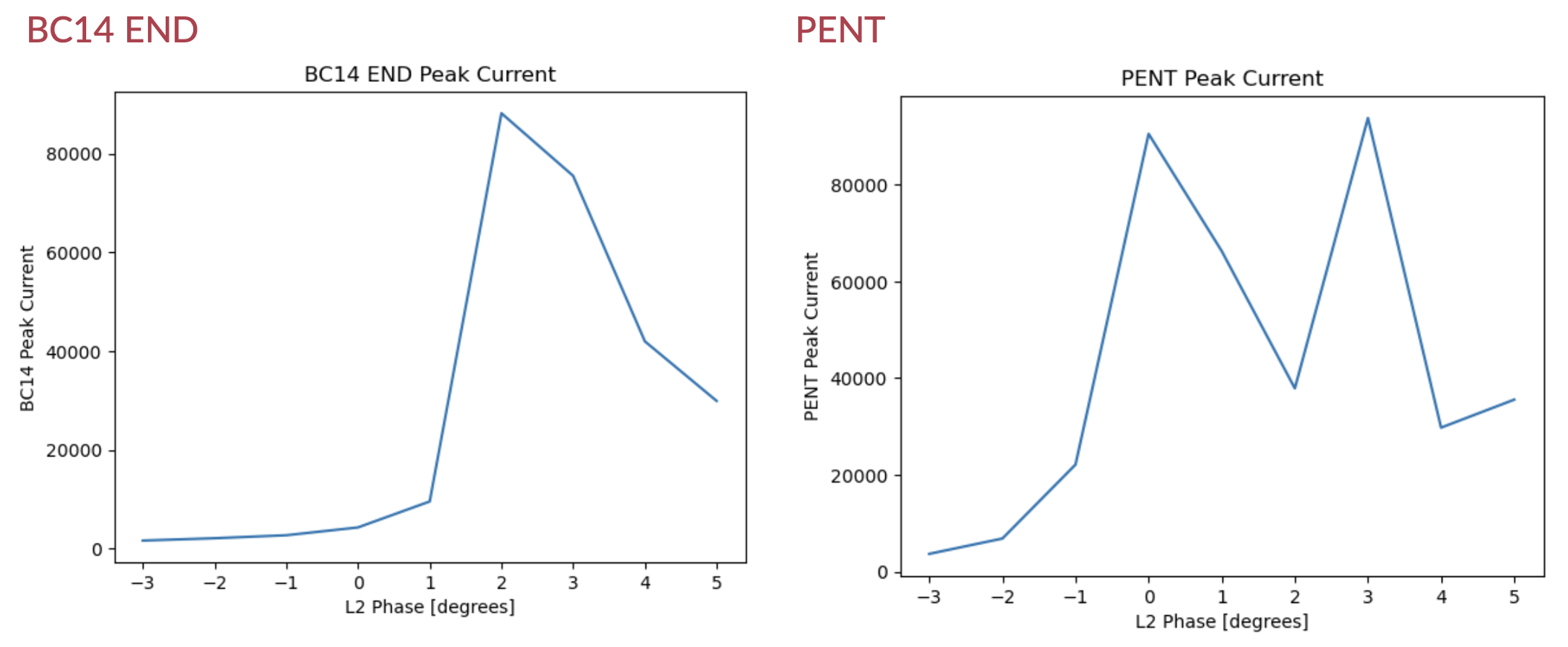}
    \caption{Peak Current Differences Based on Phase at BC14END and PENT}
    \label{fig:synthetic_reconstruction}
\end{figure}

Another important aspect of the research was the study of how the location of the beam dump affects the peak current. The beam dump is the point at which the beam is safely dissipated after it has been used. The simulations showed that the location of the beam dump could have a significant impact on the peak current, with different locations leading to variations in the final output. This finding suggests that careful consideration must be given to the placement of the beam dump in PWFA systems to ensure optimal performance. Visual differences are noted in both the current profile and longitudinal phase space of beam dumps, as shown in figure 6.

\section{Conclusion}

The research conducted at FACET-II has made significant strides in understanding and optimizing beam-plasma interactions, particularly through the detailed analysis of jitter and its impact on beam quality. The findings underscore the importance of precision in controlling various accelerator parameters, such as the phase and amplitude in the L1 and L2 cavities, to maintain high-quality beams essential for both scientific and medical applications. The use of start-to-end simulations has provided valuable insights into how small instabilities can influence critical beam characteristics, offering a pathway to more refined control mechanisms in future plasma wakefield acceleration (PWFA) systems.

Looking ahead, there are several promising avenues for further research that could build upon these findings. One key area is the continued exploration of parameter space using more sophisticated methods, such as quasi-random Gaussian selection, to identify and mitigate sources of jitter more effectively. Additionally, the calculation of correlation coefficients within a high-dimensional parameter space could provide a deeper understanding of how different variables interact, helping to pinpoint the most significant contributors to beam quality degradation.

A particularly exciting direction for future research is the integration of machine learning, specifically through the use of Bayesian Exploration techniques. Bayesian Exploration is a powerful tool for optimizing complex systems with many variables, as it allows for the systematic exploration of parameter space while incorporating prior knowledge and uncertainties. By training a neural network using Bayesian methods, researchers could create predictive models that identify optimal settings for accelerator parameters, thereby minimizing jitter and enhancing beam quality. This approach would not only improve the efficiency of PWFA systems but also accelerate the discovery process by reducing the need for exhaustive trial-and-error experiments.

The future of beam-plasma interaction research is likely to see an increasing convergence of traditional accelerator physics with advanced data science techniques. By leveraging machine learning and other computational tools, researchers can push the boundaries of what is possible with compact accelerators, paving the way for new applications in science and medicine. As the research progresses, the integration of these technologies will be critical in achieving the next generation of high-performance, cost-effective particle accelerators.




\end{document}